\documentclass[final,3p,times,twocolumn]{elsarticle}
\usepackage{lineno}
\usepackage{graphicx}
\usepackage{subcaption}
\usepackage{upgreek}
\usepackage{amsmath}
\usepackage{cuted}
\usepackage{siunitx}
\usepackage{hyperref}
\usepackage{amssymb}
\usepackage{lipsum}
\journal{Nuclear Instruments and Methods in Physics Research Section A}

\begin{document}

\begin{frontmatter}



\title{TCAD Simulations of Humidity-Induced Breakdown of Silicon Sensors}


\author[a]{I-S. Ninca} \author[a]{I. Bloch} \author[a]{B. Brüers} \author[b]{V. Fadeyev} \author[c,d]{J. Fernandez-Tejero} \author[e]{C. Jessiman} \author[e]{J. Keller} \author[e]{C. T. Klein} \author[e]{T. Koffas} \author[f]{H. M. Lacker} \author[f]{P. Li} \author[f]{C. Scharf} \author[e]{E. Staats} \author[g]{M. Ullan} \author[h]{Y. Unno}

\affiliation[a]{organization={DESY Zeuthen, Platanenallee 6, 15738 Zeuthen, Germany
}%Department and Organization
            %addressline={}, 
            %city={},
            %postcode={}, 
            %state={},
            %country={}
            }
            
\affiliation[b]{organization={Santa Cruz Institute for Particle Physics (SCIPP), University of California, Santa Cruz, CA 95064, USA
}%Department and Organization
            %addressline={}, 
            %city={},
            %postcode={}, 
            %state={},
            %country={}
            }

\affiliation[c]{organization={Department of Physics, Simon Fraser University, 8888 University Drive, Burnaby, B.C. V5A 1S6, Canada
}%Department and Organization
            %addressline={}, 
            %city={},
            %postcode={}, 
            %state={},
            %country={}
            }

\affiliation[d]{organization={TRIUMF, 4004 Wesbrook Mall, Vancouver V6T 2A3, BC, Canada
}%Department and Organization
            %addressline={}, 
            %city={},
            %postcode={}, 
            %state={},
            %country={}
            }

\affiliation[e]{organization={Physics Department, Carleton University, 1125 Colonel By Drive, Ottawa, Ontario, K1S 5B6, Canada
}%Department and Organization
            %addressline={}, 
            %city={},
            %postcode={}, 
            %state={},
            %country={}
            }

\affiliation[f]{organization={Institut für Physik, Humboldt-Universität zu Berlin, Newtonstraße 15, 12489 Berlin, Germany
}%Department and Organization
            %addressline={}, 
            %city={},
            %postcode={}, 
            %state={},
            %country={}
            }

\affiliation[g]{organization={Instituto de Microelectrónica de Barcelona (IMB-CNM), CSIC, Campus UAB-Bellaterra, 08193 Barcelona, Spain
}%Department and Organization
            %addressline={}, 
            %city={},
            %postcode={}, 
            %state={},
            %country={}
            }

\affiliation[h]{organization={Institute of Particle and Nuclear Study, High Energy Accelerator Research Organization (KEK), 1-1 Oho, Tsukuba, Ibaraki 305-0801, Japan
}%Department and Organization
            %addressline={}, 
            %city={},
            %postcode={}, 
            %state={},
            %country={}
            }

\begin{abstract}
The breakdown voltage of silicon sensors is known to be affected by the ambient humidity. To understand the sensor's humidity sensitivity, Synopsys TCAD was used to simulate n-in-p sensors for different effective relative humidities. Photon emission of hot electrons was imaged with a microscope to locate breakdown in the edge-region of the sensor. The Top-Transient Current Technique was used to measure charge transport near the surface in the breakdown region of the sensor. 
Using the measurements and simulations, the evolution of the electric field with relative humidity and the carrier densities towards breakdown in the periphery of p-bulk silicon sensors are investigated. 
\end{abstract}




\end{frontmatter}

%\tableofcontents



\section{Introduction}
\label{introduction}

The ATLAS Inner Tracker (ITk) \citep{CERN-LHCC-2017-005} will be upgraded to sustain the harsh radiation levels due to the increase in luminosity foreseen at the High-Luminosity Large Hadron Collider (HL-LHC). In the prototyping phase for the new $\text{ATLAS ITk Strip detector}$, silicon sensors showed electrical breakdown at lower bias voltages when exposed to high relative humidity (RH) of \,$\geq\,40\,\%$\,\citep{fernandez2020humidity, fernandez2023analysis}\footnote{To mitigate this issue the sensors are stored in a dry environment which corresponds to $\text{RH}\,\leq\,10\,\%$ compared to low RH. Once installed, the ITk will be in a dry environment and humidity sensitivity will not be an issue during ATLAS operation.} compared to low RH of $\leq\,10\,\%$. Although the humidity sensitivity of silicon sensors has been a known issue \citep{atalla1959stability, shockley1964mobile, kao1967high}, the mechanisms that cause early breakdown of sensors in humid conditions are not fully understood. 

In Fig.\,8 from reference \citep{fernandez2020humidity}, the humidity induced avalanche breakdown was localized in the guard ring (GR) region of the sensors. Technology Computer Aided Design (TCAD) simulations and Top-Transient Current Technique (Top-TCT) scans were performed in this region to gain insight into the physical processes triggering early breakdown due to humidity exposure. The electrical behavior of test structures was simulated using TCAD from Synopsis\,\citep{TCAD} at $\text{RH}\,=\,30\,\%\,\text{and}\,40\,\%$. TCAD simulations can give an estimate on the electric field distribution, charge transport and for how long sensors would survive in humid conditions and high bias voltage. For the first time, the Top-TCT method was used to study the charge transport in the GR region of test structures by generating localized free charge carriers near the surface with picosecond pulses of red laser light. The induced transient currents were measured as a function of RH to determine how humidity impacts the prompt current and charge profile. The prompt current can potentially provide information about the electric field. 

\section{TCAD Simulations}

\subsection{Breakdown imaging}

\begin{figure}[!ht]
	\centering 
	\includegraphics[width=0.25\textwidth]{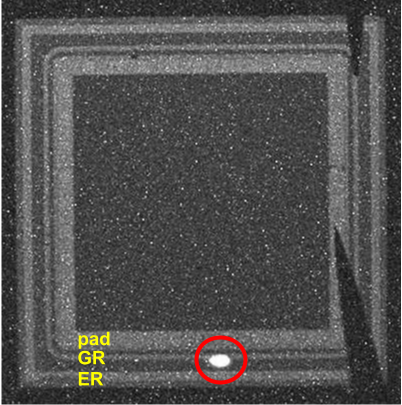}	
	\caption{Microscope picture of an n-in-p silicon diode operated in avalanche breakdown. Photon emission from hot electrons in the breakdown region is visible as a bright spot in the bottom of the image circled in red. The measurement was performed at a relative humidity (RH) of $\approx$50\,\%. The breakdown current between the backside and the GR was $\mathrm{\mathcal{O}(1)~\si{\micro\ampere}}$.} 
	\label{hotspot}%
\end{figure}

To cross check the location of the breakdown region, a consumer CMOS camera sensitive in the near infrared spectrum \citep{Camera} was used to image test structures. Fig.\,\ref{hotspot} shows an example $4\times4$\,mm$^2$ n-in-p diode, biased at $\mathrm{-~750~V}$ in electrical breakdown. The outermost aluminum ring is called the edge ring (ER), then there is the guard ring (GR) and finally the pad metal. The bias voltage was applied on the backside, which has a conductive coupling to the ER via the undepleted region along the dicing edge. The GR and pad were kept at ground. The bright spot is Bremsstrahlung from hot electrons\,\citep{akil1999multimechanism, bude1992hot} accelerated in a high electric field between the ER and the GR.

\subsection{Device under test}

For this study, an $\mathrm{8}$ x $\mathrm{8~mm}^2$ n-in-p diode called the monitor diode (MD8) was investigated. This type of diode comes from the same ATLAS18 ITk wafer as the strip sensor\,\citep{unno2023specifications}. The MD8’s geometry is similar to the diode illustrated in Fig.\,\ref{hotspot} but it has a special feature of an additional p-stop between the GR and the pad metal. The active thickness calculated from capacitance-voltage measurements\,\citep{Riemer:471495} is $\mathrm{295~\upmu m}$ with a full depletion voltage of $\approx-280$\,V and the effective p-bulk doping is $\approx4.2\cdot10^{12}$\,cm\,$^{-3}$.

\subsection{TCAD geometry implementation}

\begin{figure}[!ht]
	\centering 	\includegraphics[width=0.4\textwidth, angle=0]{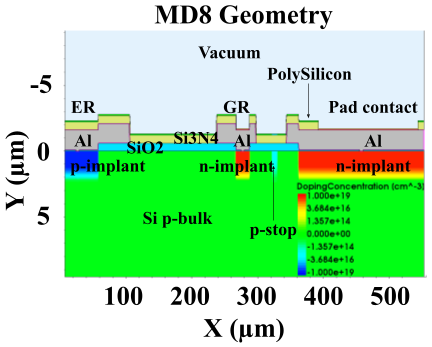}	
	\caption{View of the diode geometry used in TCAD. The backside p-implant (not shown) is at $\text{Y}\,=\,297.2\,\upmu$m and extends along $\text{X}$.} 
	\label{Geo}%
\end{figure}

The Structure Device Editor (SDE) from Synopsys was used to implement the MD8 geometry for the TCAD simulations using representative parameters. The cross-section of the test structure is shown in Fig.\,\ref{Geo} with a focus on the top edge of the diode. There is a p-implant underneath the ER, while the GR and pad have n-implants. The absolute effective doping concentrations of both p- and n-implants are $\mathrm{10^{\text{19}}\,cm^{-3}}$. There is a $\mathrm{0.6\,\upmu m\,SiO_2}$ layer in direct contact with the silicon bulk. A $\mathrm{0.6\,\upmu m\,Si_3N_4}$ layer is placed on top of the $\mathrm{SiO_2}$ and the electrodes. The concentration of fixed oxide charges at the interface between the silicon bulk and the $\mathrm{SiO_2}$ is $\mathrm{\approx\,10^{11}\,cm^{-2}}$\,\citep{ullan2020quality}. The values for the parameters implemented are listed in Table\,\ref{Par}.

\begin{table}[!ht]
  \centering
  \caption{TCAD Parameter Values}
  \begin{tabular}{|c|c|c|}
    \hline
    Parameter & Value & Unit \\
    \hline
    Si p-bulk thickness & $295$ & $\upmu$m \\
    Si p-bulk doping & $4.2 \cdot 10^{12}$ & cm$^{-3}$ \\
    p- and n-implant doping & $10^{19}$ & cm$^{-3}$ \\
    %n-implant doping & $1 \cdot 10^{19}$ & cm$^{-3}$ \\
    p-stop doping & $10^{16}$ & cm$^{-3}$ \\
    Fixed oxide charge & $10^{11}$ & cm$^{-2}$ \\
    SiO$_2$/Si$_3$N$_4$ thicknesses & $0.6$ & $\upmu$m \\
    %Si$_3$N$_4$ thickness & $0.6$ & $\upmu$m \\
    %Poly-Si thickness & $0.1$ & $\upmu$m \\
    %Al electrodes thickness & $1.5$ & $\upmu$m \\    
    \hline
  \end{tabular}
  \label{Par}%
\end{table}

%To model the effects of humidity on the surface of the test structure, a 0.1\,$\upmu$m polysilicon layer was added on top of the passivation which acts as a resistive layer. Polysilicon is not on the physical device, but this method has been previously proven to work in \citep{schwandt2017surface}. The polysilicon is directly connected to the GR and pad, but it is not in direct contact with the ER since the ER only has passivation openings in the corners of the diode. 
In the presence of humidity, the sheet resistance, $\text{R}_{\square}$, of the passivation surface decreases with increasing the RH as measured in\,\citep{7431261}. This is driven by the redistribution of ions on the surface and the RH affects the mobility of these surface ions. In Synopsis TCAD ion motion cannot be simulated directly. Instead, it is modeled with a 0.1\,$\upmu$m polysilicon layer added on top of the passivation, assigning very low mobilities to the electrons and holes to match the measured $\text{R}_{\square}$ caused by ion motion. The polysilicon is directly connected to the GR and pad, but it is not in direct contact with the ER since the ER only has passivation openings in the corners of the diode. The coupling is important because electrons and holes can move from the polysilicon layer into the aluminum electrodes. Ions do not behave in the same way, although they could potentially be neutralized at the electrodes. The charge carriers mobilities were calculated based on the $\text{R}_{\square}$ measurements presented in \citep{schwandt2017surface} and the values used are recorded in Table\,\ref{RHValues}. 

\begin{table}[!ht]
  \centering
  \caption{TCAD Mobility Values}
  \begin{tabular}{|c|c|c|}
    \hline
    RH & $\upmu_\text{e,h}$ & Unit \\
    \hline
    30 \% & $9.86\cdot10^{-4}$ & $\text{cm}^2/\text{Vs}$ \\
    40 \% & $6.71\cdot10^{-3}$  & $\text{cm}^2/\text{Vs}$ \\
        \hline
  \end{tabular}
  \label{RHValues}%
\end{table}

The carrier mobilities in silicon were described by the Canali model\,\citep{canali1975electron} for the high-field saturation and the University of Bologna model\,\citep{reggiani2002electron} for the doping dependence. Shockley–Read–Hall (SRH) recombination with the lifetime $\uptau=10^{-2}$\,s and Auger recombination were used. Charge carrier multiplication was simulated using the van Overstraeten-de Man model\,\citep{van1970measurement}, which relies on the Chynoweth law\,\citep{chynoweth1958ionization}.

\subsection{TCAD Results}

\begin{figure}[!ht]
	\centering 	\includegraphics[width=0.45\textwidth, angle=0]{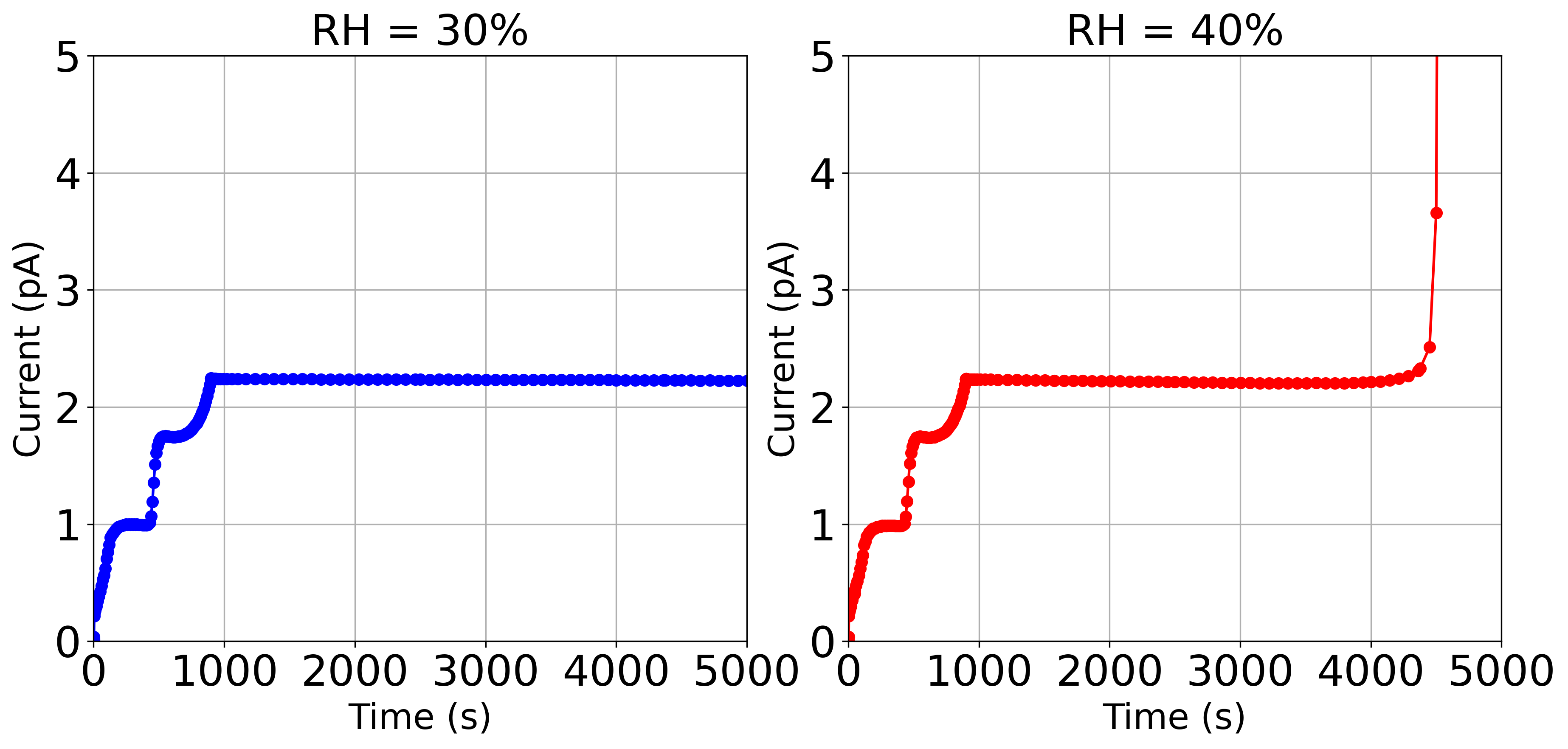}	
	\caption{Simulated GR current for $\text{RH}\,=\,30\,\%\,\text{and}\,40\,\%$ at $\text{V}_{\text{const.}}\,=\,-\,900\,\text{V}$. After ramping, the leakage current is of the order of $10^{-12}\,\text{A}$ for the entire duration of the simulation at $\text{RH}\,=\,30\,\%$. At $\text{RH}\,=\,40\,\%$, it remains at this level until electrical breakdown starts at $\text{t}\,=\,4512\,\text{s}$.}
	\label{IV}%
\end{figure}

In TCAD, the bias voltage was applied to the backside, which is electrically connected to the ER, starting from 0 V and increasing in steps of -\,10\,V every 10\,s. The GR and pad were kept at 0\,V. After $\mathrm{900\,s}$, the bias voltage\footnote{Fig.\,2 in \citep{fernandez2023analysis} shows that the breakdown voltage for an initial current-voltage measurement of a prototype ATLAS17LS full-size sensor exposed to high humidity (RH\,=\,41\,\%) was less than -\,800\,V. To reduce the simulation run time, a bias voltage of -\,900\,V was considered sufficient to examine the effects of humidity on the monitoring diode.} was kept constant at $\mathrm{-\,900\,V}$. Fig.\,\ref{IV} shows the current evolution with time at two RH values (30\,\% and 40\,\%). For $\text{RH}\,=\,30\,\%$, the leakage current is of the order of $10^{-12}~\text{A}$ after the ramping period and no electrical breakdown occurred during the simulation time (11700\,s). For $\text{RH}\,=\,40\,\%$, the leakage current starts to increase at $\text{t}\,=\,4512\,\text{s}$. To understand what drives the breakdown of sensors in humid conditions and high bias voltage, the electric field and charge carrier densities are presented immediately after ramping and after waiting a defined time at constant bias.

\begin{figure}[!ht]
	\centering 	\includegraphics[width=0.49\textwidth, angle=0]{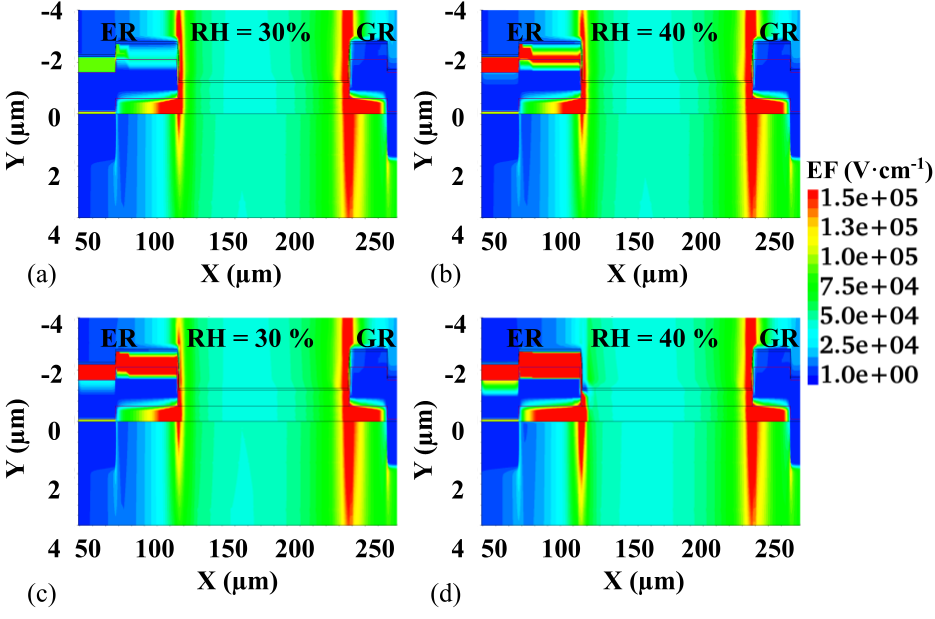}	
	\caption{The absolute electric fields at $\text{RH}\,=\,30\,\%\,\text{and}\,40\,\%$ at $\text{t}\,=\,900\,\text{s}$ are shown in a)\,and\,b). The absolute electric fields at $\text{RH}\,=\,30\,\%\,\text{and}\,40\,\%$ at $\text{t}\,=\,4550\,\text{s}$ are shown in c)\,and\,d). } 
	\label{EF-900-4550}%
\end{figure}

The 2D representations of the absolute electric field distributions at $\text{RH}\,=\,30\,\%$ and $\text{RH}\,=\,40\,\%$ for $\text{t}\,=\,900\,\text{s}$ are showcased in Fig.\,\ref{EF-900-4550}\,a)\,and\,b). The color map was manually adjusted: the minimum  was set to $0\,\text{V}\,\cdot\,\text{cm}^{-1}$, and the maximum was set to $1.5\cdot10^5\,\text{V}\cdot\text{cm}^{-1}$. All values $\geq\,1.5\cdot10^5\,\text{V}\cdot\text{cm}^{-1}$ are represented by the shade of red, which is considered to be the maximum of the plot range. For both RH values, two high field peak regions ($\geq\,1.5 \cdot 10^5\,\text{V}\cdot\,\text{cm}^{-1}$) are formed near the ER and GR. Fig.\,\ref{EF-900-4550}\,c)\,and\,d) show the electric fields at $\text{t}\,=\,4550\,\text{s}$ and $\text{V}_\text{bias}\,=\,-900\,\text{V}$. Fig.\,\ref{EF-900-4550}\,d) illustrates the high-field regions visible inside the $\text{SiO}_2$ layer near the ER and GR extending laterally.

\begin{figure}[!ht]
	\centering 
	\includegraphics[width=0.349\textwidth, angle=0]{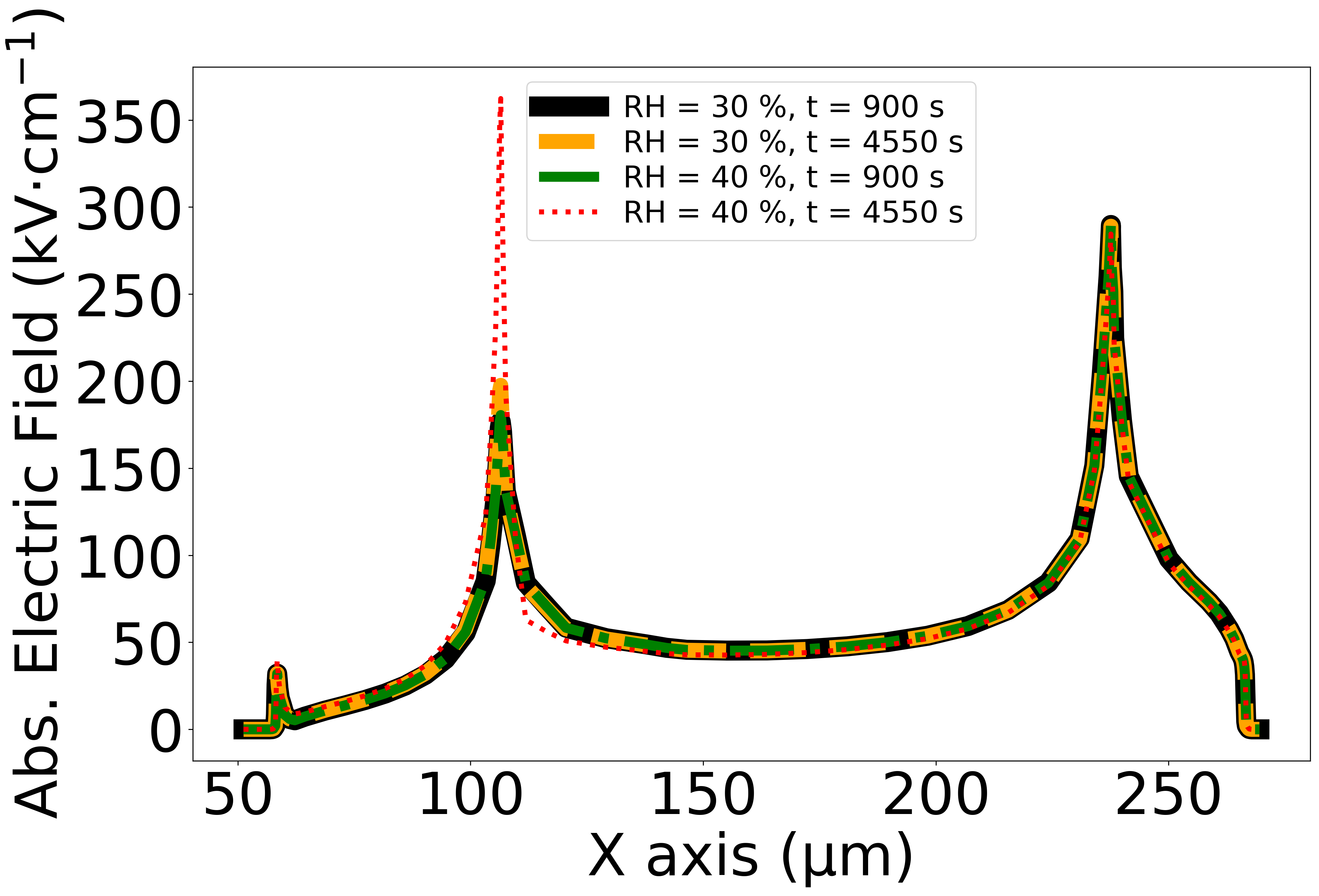}	
	\caption{Values of the absolute electric field along $\text{X}$ axis at $\text{Y}\,=\,100\,\text{nm}$. The ER is between $0\,\upmu \text{m}\,\text{and}\,106\,\upmu \text{m}$, and the GR is between $237\,\upmu \text{m}\,\text{and}\,298\,\upmu \text{m}$. } 
	\label{Cutlines}%
\end{figure}

The absolute electric field values at $\text{Y}~=~100~\text{nm}$ along the X-axis are presented in Fig.\,\ref{Cutlines} where the maximum of the amplitude of the absolute electric field has not been adjusted. At $\text{RH}\,=\,30\,\%$, the high field peak underneath the ER increases from $1.8\cdot10^5\,\text{V}\cdot\text{cm}^{-1}$ to $2.1\cdot10^5\,\text{V}\cdot\text{cm}^{-1}$ after waiting some time. At $\text{RH}\,=\,40\,\%$, the high field peak underneath the ER increases from $1.8\cdot10^5\,\text{V}\cdot\text{cm}^{-1}$ to $\geq\,3.5\cdot10^5\,\text{V}\cdot\text{cm}^{-1}$. At the same time, the field peak underneath the GR stays the same for both RH values from $900\,\text{s}$ until $4550\,\text{s}$.

\begin{figure}[!ht]
	\centering 
	\includegraphics[width=0.49\textwidth]{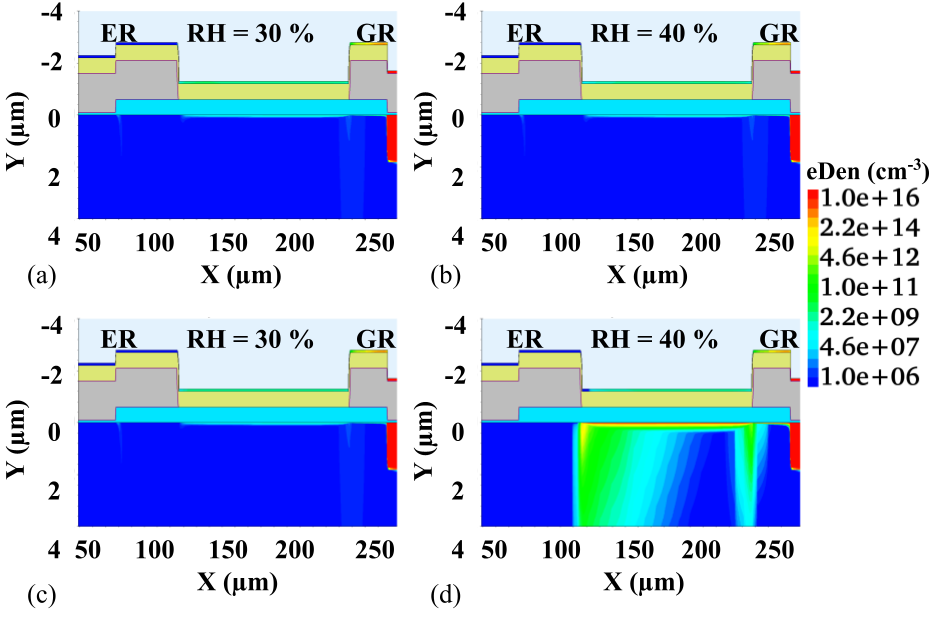}	
	\caption{The electron density at $\text{RH}\,=\,30\,\%\,\text{and}\,40\,\%$ at $\text{t}\,=\,900\,\text{s}$ are shown in a)\,and\,b). The electron density at $\text{RH}\,=\,30\,\%\,\text{and}\,40\,\%$ at $\text{t}\,=\,4550\,\text{s}$ are shown in c)\,and\,d).} 
	\label{eDen-900-4550}%
\end{figure}

The electron densities immediately after ramping ($\text{t}\,=\,900\,\text{s}$) are shown in Fig.\,\ref{eDen-900-4550}\,a)\,and\,b). Due to the fixed positive oxide charge resulting from the fabrication process, a conducting electron inversion layer is present at the Si-SiO$_2$ interface before a bias voltage is applied. The electron inversion layer is dispersed by the electric field when the interface is depleted and the remaining $10^7\,\text{cm}^{-3}$ electron concentration at the interface is low after ramping to $\mathrm{-\,900\,V}$, as seen in Fig.\,\ref{eDen-900-4550}\,a)\,and\,b). The increase of the current between $4512\,\text{s}\,\text{and}\,4550\,\text{s}$ at $\text{RH}\,=\,40\,\%$ coincides with an increase of the electron concentration in the inversion layer from $10^7\,\text{cm}^{-3}$ up to $10^{16}\,\text{cm}^{-3}$, as visible in Fig.\,\ref{eDen-900-4550}\,d). 

\begin{figure}[!ht]
	\centering 	\includegraphics[width=0.49\textwidth]{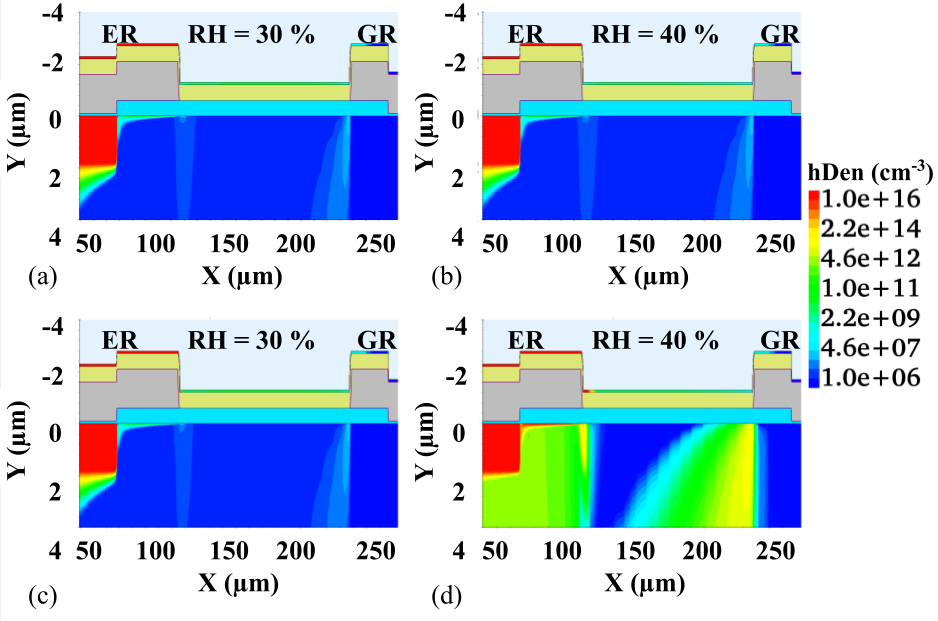}	
	\caption{The hole density at $\text{RH}\,=\,30\,\%\,\text{and}\,40\,\%$ at $\text{t}\,=\,900\,\text{s}$ are shown in a)\,and\,b). The hole density at $\text{RH}\,=\,30\,\%\,\text{and}\,40\,\%$ at $\text{t}\,=\,4550\,\text{s}$ are shown in c)\,and\,d). } 
	\label{hDen-900-4550}%
\end{figure}

The evolution of the concentration of free holes in the bulk is comparable to the electron concentration, as shown in Fig.\,\ref{hDen-900-4550}\,a)\,and\,b). A high density of holes ($10^{16}\,\text{cm}^{-3}$) is present in the polysilicon layer on top of the ER after the ramping of the device has finished. It is important to note that the polysilicon layer is capacitively coupled to the ER, while it is in direct electrical contact with the GR and the pad. Because there are no passivation openings for the ER, charges accumulate on top of this electrode. Over time, at  $\text{RH}\,=\,40\,\%$ a high concentration of holes in the polysilicon layer moves laterally towards the GR, visible at $\text{X}\approx110\,\upmu$m in Fig.\,\ref{hDen-900-4550}\,d). The observed charging up of the passivation surface is in agreement with \citep{atalla1959stability}. This effect was considered to be the leading factor driving the breakdown because the hole accumulation on top of the ER changes the electrical properties of the sensor. 

%$>\,10^5\,\text{V}\cdot\text{cm}^{-1}$ \citep{maes1990impact}. 

\begin{figure}[!ht]
	\centering 	\includegraphics[width=0.49\textwidth]{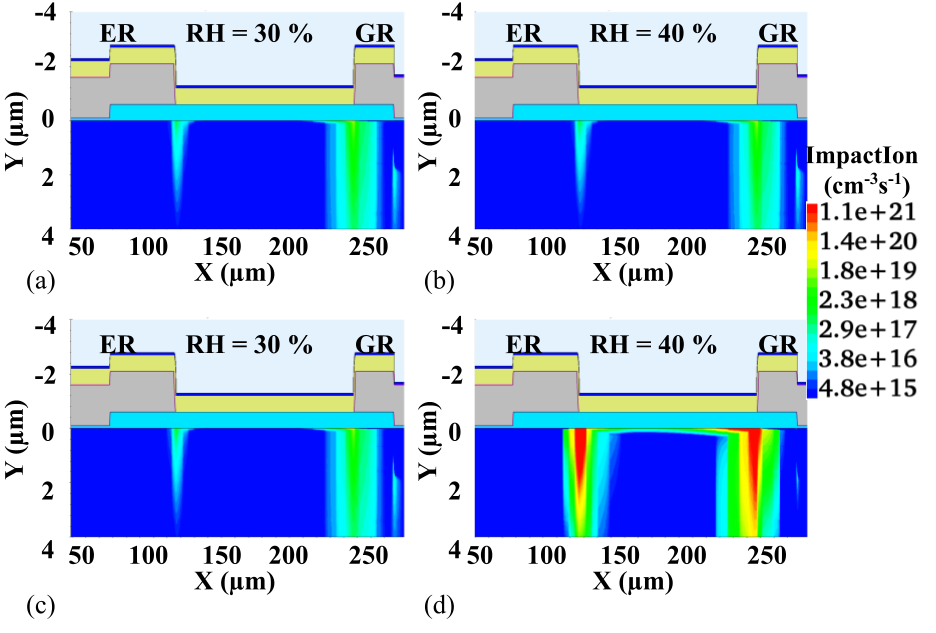}	
	\caption{The electron-hole pair production due to impact ionization at $\text{RH}\,=\,30\,\%\,\text{and}\,40\,\%$ at $\text{t}\,=\,900\,\text{s}$ are shown in a)\,and\,b). The electron-hole pair production due to impact ionization at $\text{RH}\,=\,30\,\%\,\text{and}\,40\,\%$ at $\text{t}\,=\,4550\,\text{s}$ are shown in c)\,and\,d). For comparison, the bulk generation rates is $\text{U}\,=\,4.78\cdot10^{15}\,\text{cm}^{-3}\,\text{s}^{-1}$.} 
	\label{ImpactIon-900-4550}%
\end{figure}

Electron-hole pair production due to impact ionization requires a certain electric field strength. After ramping, there is an impact ionisation region underneath the metal overhang of the ER and GR as illustrated in Fig.\,\ref{ImpactIon-900-4550}\,a)\,and\,b). Fig.\,\ref{ImpactIon-900-4550}\,d) shows an increase in electron-hole pair production due to impact ionization, reaching $10^{21}\,\text{cm}^{-3}\text{s}^{-1}$. This is correlated to the previous results in Fig.\ref{eDen-900-4550}\,d) and \ref{hDen-900-4550}\,d) where electron-hole pairs are generated by impact ionization in the high field regions at the ER and GR (see Fig.\,\ref{EF-900-4550}\,d).

\section{Top-TCT Measurements}

\begin{figure}[!ht]
	\centering 
	\includegraphics[width=0.35\textwidth, angle=0]{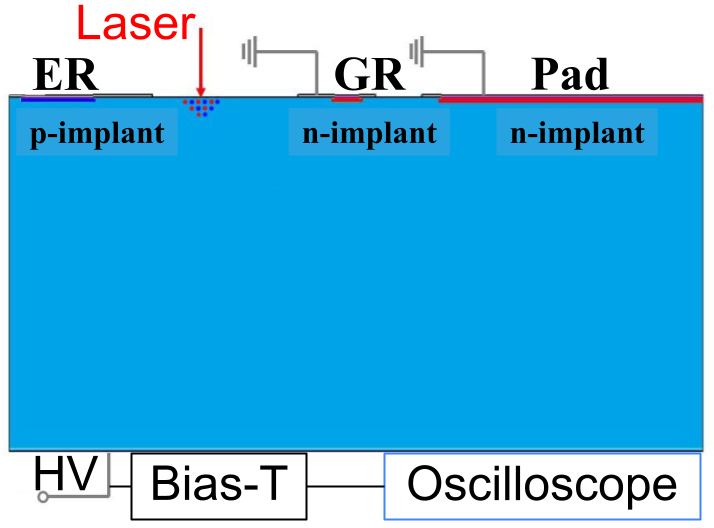}	
	\caption{Top-TCT sketch showing electron-hole pair generation by a $\mathrm{660~nm}$ laser pulse between the ER and the GR. During a TCT measurement, the laser beam is moved along the surface of the diode and the induced transients are recorded for each position.} 
	\label{TCT-setup}%
\end{figure}

The Top-TCT set-up from Particulars \citep{Particulars} was used to study the charge transport in the region between the ER and the pad of the MD8 diode. Laser pulses of $\mathrm{660~nm}$ wavelength were focused at the sensor's surface with a full width at half maximum of the laser beam $\text{FWHM}=7.33\,\upmu$m$\,\pm\,0.16\,\upmu$m. The charge injected corresponds to an equivalent of 12 MIPs creating $3.6\cdot10^5$ electron-hole pairs. The backside was connected to high voltage (HV) and read out via a Bias-T. The ER was also at HV through a conductive channel at the dicing edge to the backside. The GR and pad were connected to ground\footnote{Normally, the GR is floating to maximize the breakdown voltage.} as depicted in Fig.\,\ref{TCT-setup}. 

The laser was moved across the diode’s surface and the resulting transient current from the drifting charge carriers was recorded at the backside, which has a conductive connection to the ER. A Rohde and Schwarz RTO1024 oscilloscope with a sampling rate of $\mathrm{10\,Gsample/s}$ and a bandwidth of $\mathrm{2\,GHz}$ was used to acquire the current transients averaged over 1000 laser pulses. The humidity inside the TCT set-up was monitored using a humidity sensor placed in the vicinity of the diode. The humidity was controlled by flushing the TCT set-up with dry air until the desired RH value was reached. During the measurements, the RH was stable within $\mathrm{\pm\,3\,\%}$. 

Prior to TCT scans, the MD8 diodes were measured in a humidity-controlled probe station. For each RH value (10\,\%, 20\,\%, 30\,\%, 40\,\%), current-voltage (IV) measurements were performed. None of the tested MD8 diodes experienced break down and hot-electron emission could not be recorded. Previous studies have shown that small structures like mini sensors \citep{unno2023specifications} or diodes show less humidity sensitivity than large ones \citep{fernandez2023analysis}, such as the ATLAS ITk strip sensors. Top-TCT measurements were performed for the exact same RH value. The humidity was ramped up throughout the tests, starting at $\mathrm{RH\,=\,10\,\%}$ to avoid hysteresis effects. These can occur when the sensor is biased at high humidity and the humidity is subsequently lowered, freezing a possible charge at the surface\,\citep{shockley1964mobile}. The results shown are from a single diode, MD8 32418-14. For all experimental results, the applied bias voltage was $\mathrm{-\,900\,V}$. The data acquisition time was approx. $600\,\text{s}$.

The transient current induced by the drift of charge carriers at the collection electrode is given by the Shockley-Ramo theorem\,\citep{ramo1939currents}:
\begin{equation}
    \text{I} \ = \ \text{e} \cdot \upmu_\text{e,h}(\vec{\text{E}}) \cdot \vec{\text{E}} \cdot \vec{\text{E}_\text{w}}
    \label{Curr-eq}%
\end{equation}
where e is the elementary charge, $\upmu_{\text{e,h}}$ is the mobility, $\vec{\text{E}}$ is the electric field and $\vec{\text{E}}_{\text{w}}$ is the weighting field. Equation\,\ref{Curr-eq} is representative for the current induced by a single charge carrier and it is only used to understand the Top-TCT measurements. 

%The measured induced current is the sum of both electrons and holes induced currents: $\text{I}_\text{measured}(\text{t})=\text{I}_\text{e}(\text{t})+\text{I}_\text{h}(\text{t})$. 
%Charge carriers drifting towards the measuring electrode(s) will undergo the effects of the weighting and electric fields which can induce very different currents that are superimposed. 

\subsection{Top-TCT Results}

\begin{figure}[!ht]
	\centering 
	\includegraphics[width=0.4\textwidth, angle=0]{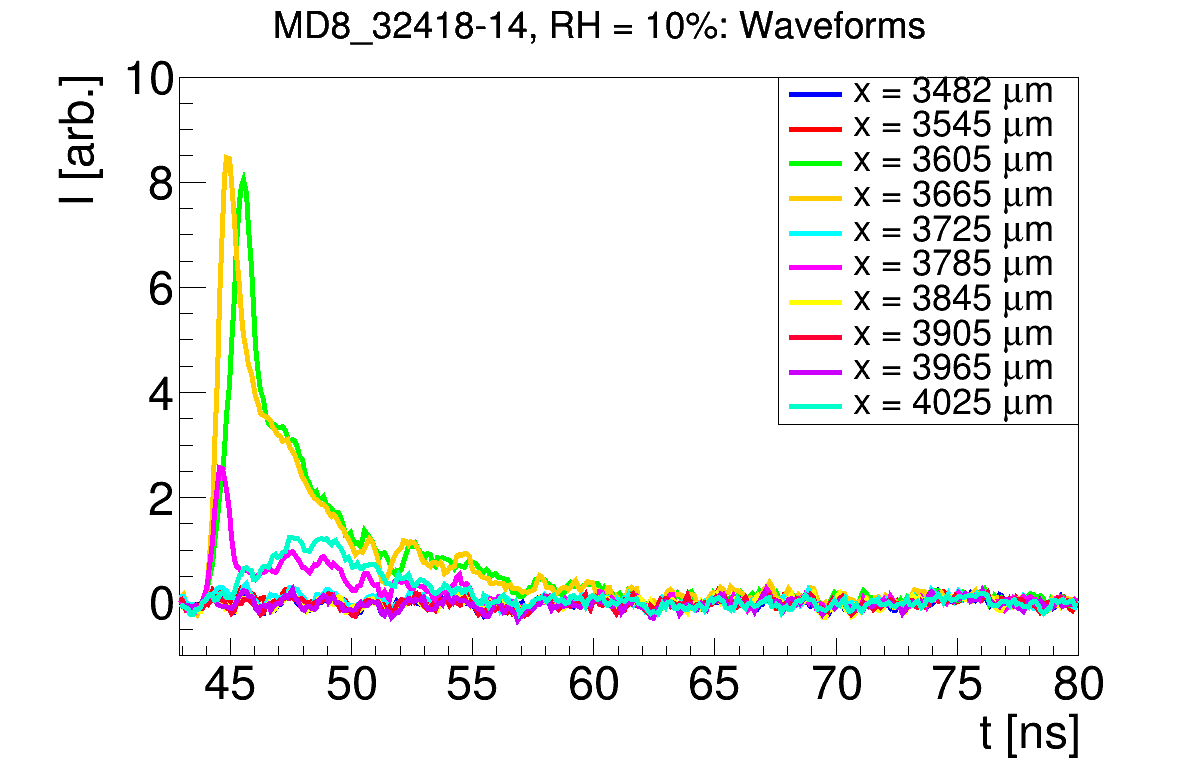}	
	\caption{Measured transient currents at different positions in the edge-region of the diode for $10\,\%\,\text{RH}$ and $\text{V}_\text{bias}\,=\,-\,900\,\text{V}$. The position at $\text{X}\,=\,3482\,\upmu \text{m}$ is close to the ER while the position at $X=4025\,\upmu \text{m}$ is close to the pad. The position of the GR is around $X=3700\,\upmu \text{m}$.} 
	\label{Waveforms}%
\end{figure}

\begin{figure}[!ht]
	\centering 
	\includegraphics[width=0.4\textwidth, angle=0]{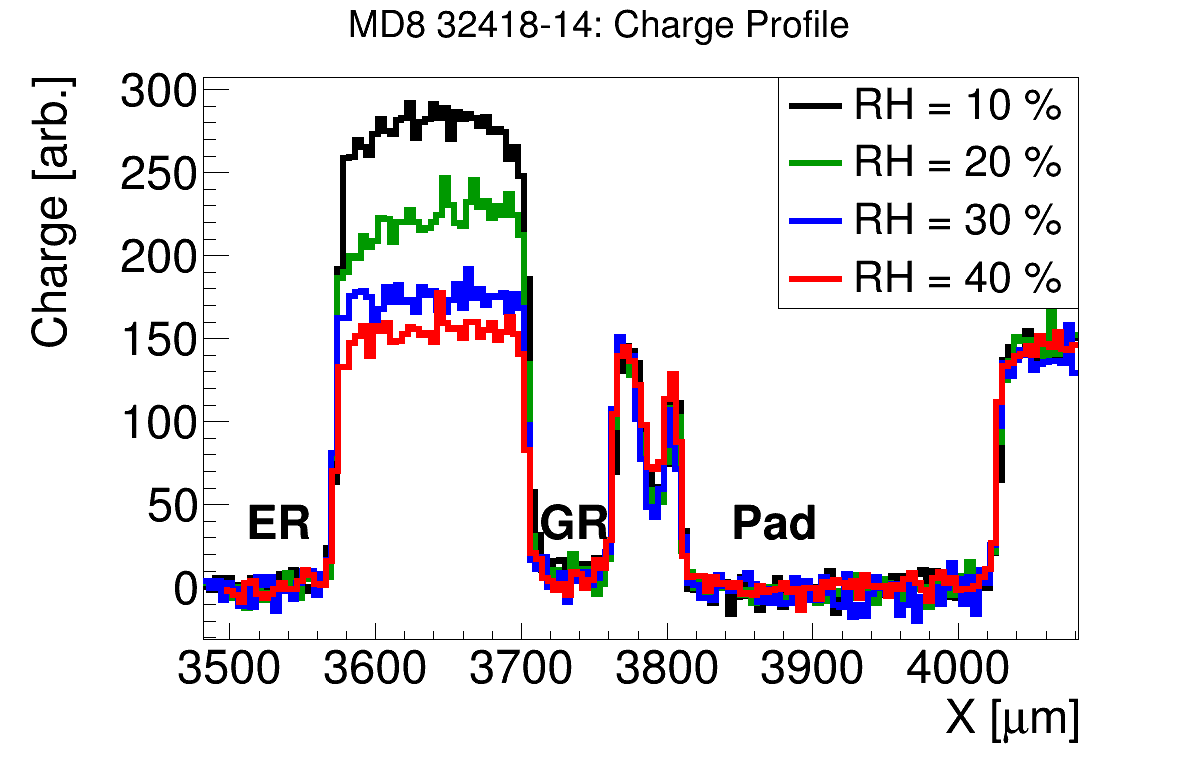}	
	\caption{Charge profiles between the ER and pad metalization for different values of RH and $\text{V}_{\text{bias}}\,=\,-900$\,V.} 
\label{Charge}%
\end{figure}

Fig.\,\ref{Waveforms} shows the transient currents at $\text{RH}\,=\,10\,\%$ and multiple positions along a straight line between the ER and the pad metal. Integrating the transient currents expressed in Eq.\,\ref{Curr-eq} for 43\,ns $\leq \text{t} \leq$ 80\,ns, the total collected charge can be obtained as a function of the laser position, shown in the charge profiles in Fig.\,\ref{Charge}. Between the ER and GR, the charge profile decreases with increasing the RH. At the same time, the collected charge between the GR and the pad does not change. The high collected charge between the ER and the GR for low humidity is caused by charge multiplication in the high field at the edge of the GR. 

It is not yet understood why charge multiplication decreases with the RH. Simulations of the injected charge in humid conditions are planned to understand this phenomena. 

\begin{figure}[!ht]
	\centering 
	\includegraphics[width=0.4\textwidth, angle=0]{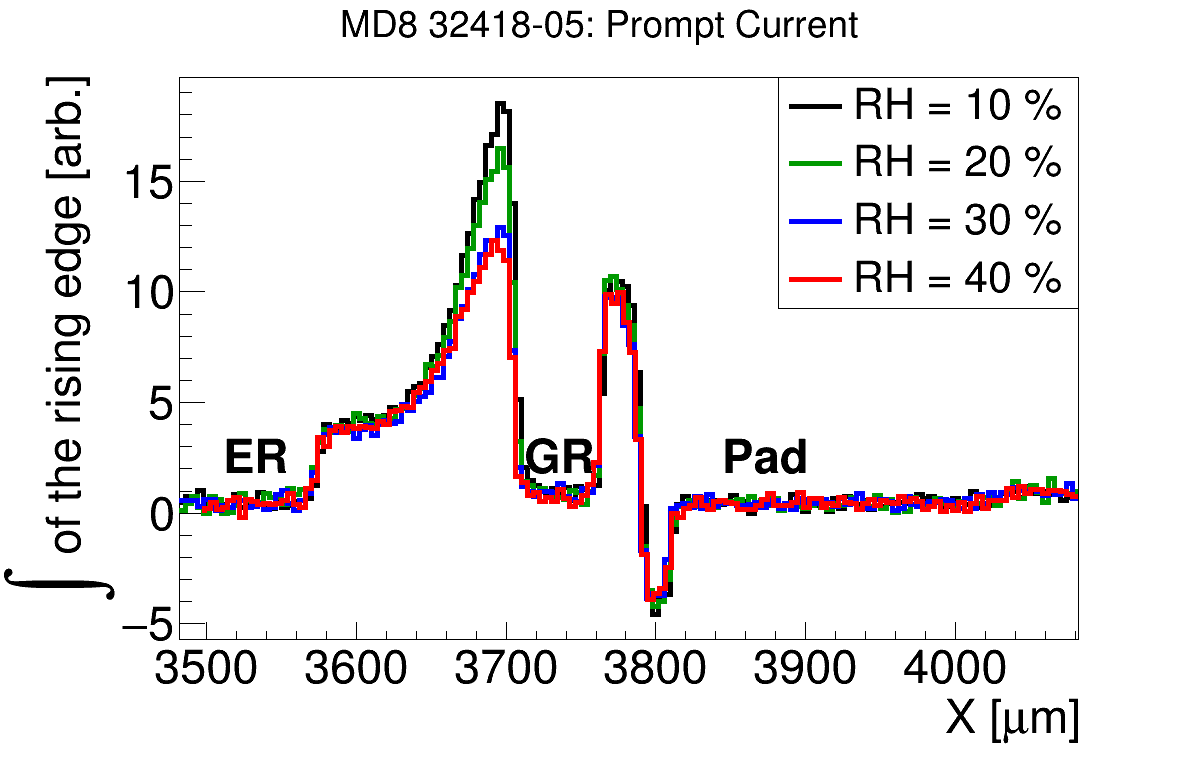}	
	\caption{Prompt currents between the ER and pad metalization for different RH and $\text{V}_{\text{bias}}\,=\,-900$\,V.} 
	\label{Velocity}%
\end{figure}

Using a short integration window at the beginning of the current transient, 44.0\,ns $\leq t \leq$ 44.4\,ns, the prompt current profiles\,\citep{kramberger2010investigation} are obtained as shown in Fig.\ref{Velocity}. The prompt current is the first step towards deriving the electric field. Currently, the prompt current provides an indication of the electric field behavior. The maximum prompt current occurs near the GR and decreases as the laser moves away from this electrode, confirming the presence of a high electric field peak near the GR. This cross-check between the simulated electric field distributions and the prompt current from TCT is important because it explains how charges could be accelerated in this region and undergo charge multiplication leading to avalanche breakdown in humid conditions. Additionally, the maximum amplitude of the prompt current decreases with increasing RH because the prompt current is also affected by charge multiplication, as seen from the charge profiles in Fig.\ref{Charge}.

\section{Summary and discussion}

Top-TCT measurements and TCAD simulations of silicon sensors with a focus on the guard ring region have been performed at a bias voltage of $-900\,\text{V}$ to study humidity-related breakdown. Electrical breakdown is observed in the TCAD simulation at $\text{RH}\,=\,40\,\%$ after waiting for some time ($\text{t}\,=\,4512\,\text{s}$) at a constant bias voltage. No electrical breakdown is observed in TCAD for $\text{RH}\,\leq\,30\,\%$ (here, only simulation results for $\text{RH}\,\geq\,30\,\%$ are presented). During the TCT measurements, no breakdown was observed for the data acquisition time ($\text{t}\,=\,600\,\text{s}$) which is in agreement with the TCAD simulations. %In the test structure used for TCT measurements, no reproducible breakdown stable enough to perform a long term measurement was observed \footnote{Smaller devices i.e.: MD8s or MD4s have a stable electrical behavior for $\text{V}_\text{bias} \leq - 900 \ \text{V}$ also observed in \citep{fernandez2020humidity}, however if $\text{V}_\text{bias} \geq - 900 \ \text{V}$, the device can breakdown permanently}. 
The total collected charge in TCT measurements between the guard ring and the edge was observed to decrease with rising relative humidity. This behavior is not yet understood; therefore, simulating the charge injection in TCAD is planned to verify if this dependency can be reproduced and to understand the underlying causes.
The maximum prompt current location observed in the measurement coincides with an electric field peak observed in the TCAD simulation which is sufficiently high ($\geq\,3\cdot 10^5\,\text{V}\cdot\text{cm}^{-1}$) to induce charge multiplication of electrons. 
The development of the field region at the ER seems to be caused by a positive charging of the surface of the passivation at the ER. This is considered to be the principal cause for humidity-related avalanche breakdown in the guard ring region for the p-bulk sensors at high bias investigated here, assuming the resistive polysilicon layer on the passivation is directly coupled to the GR. 
Further measurements in breakdown conditions as well as further simulation studies on the influence of the coupling of the polysilicon layer and the fixed oxide charge density are planned.

\section*{Acknowledgments}

The work at CNM is part of the Spanish R\&D grant PID2021-126327OB-C22, funded by MCIN/ AEI/10.13039/501100011033 / FEDER, UE.
The work at SCIPP was supported by the US Department of Energy, grant DE-SC0010107.
The work at SFU, TRIUMF and Carleton University was supported by the Canada Foundation for Innovation and the Natural Sciences and Engineering Research Council of Canada

\section*{Declaration of AI and AI-assisted technologies in the writing process}

During the preparation of this work the author(s) used ChatGPT 3.5 in order to improve the clarity of the language. After using this tool/service, the author(s) reviewed and edited the content as needed and take(s) full responsibility for the content of the publication.

\section*{Copyright}
Copyright 2024 CERN for the benefit of the ATLAS Collaboration. Reproduction of this article or parts of it is allowed as specified in the CC-BY-4.0 license.

%%\bibliographystyle{elsarticle-harv} 
%%\bibliographystyle{elsarticle-num} 
%%\bibliography{Bibliography}



%% For example:

%% \bibitem[Aladro et al.(2015)]{Aladro15} Aladro, R., Martín, S., Riquelme, D., et al. 2015, \aas, 579, A101

%%\end{thebibliography}

\end{document}